\documentclass[nobibnotes, twocolumn, aps, superscriptaddress]{revtex4-2}
\usepackage[utf8]{inputenc}
\usepackage[letterpaper, margin=2.54cm]{geometry}
\usepackage{graphicx}
\usepackage{dcolumn}
\usepackage{bm}
\usepackage{amsmath}
\usepackage[english]{babel}
\usepackage[T1]{fontenc}
\usepackage{verbatim}
\usepackage{physics}
\usepackage{float}
\usepackage{xcolor}
\usepackage{siunitx}
\usepackage{booktabs}
\usepackage{amssymb}
\usepackage{textcomp}
\usepackage[hidelinks]{hyperref}
\usepackage{xr}

\makeatletter
\newcommand*{\addFileDependency}[1]{
\typeout{(#1)}
\@addtofilelist{#1}
\IfFileExists{#1}{}{\typeout{No file #1.}}
}\makeatother

\begin{document}
\title{Capacitive crosstalk in gate-based dispersive sensing of spin qubits}
\author{Eoin G. Kelly}
\author{Alexei Orekhov}
\author{Nico Hendrickx}
\author{Matthias Mergenthaler}
\author{Felix Schupp}
\author{Stephan Paredes}
\affiliation{IBM Research -- Zurich, Säumerstrasse 4, CH-8803 Rüschlikon, Switzerland}
\author{Rafael S. Eggli}
\author{Andreas V. Kuhlmann}
\affiliation{Department of Physics, University of Basel, Klingelbergstrasse 82, CH-4056 Basel, Switzerland}
\author{Patrick Harvey-Collard}
\author{Andreas Fuhrer}
\author{Gian Salis}
\email{gsa@zurich.ibm.com}
\affiliation{IBM Research -- Zurich, Säumerstrasse 4, CH-8803 Rüschlikon, Switzerland}

\date{\today}

\begin{abstract}

In gate-based dispersive sensing, the response of a resonator attached to a quantum dot gate is detected by a reflected radio-frequency signal. This enables fast readout of spin qubits and tune up of arrays of quantum dots, but comes at the expense of increased susceptibility to crosstalk, as the resonator can amplify spurious signals and induce fluctuations in the quantum dot potential. We attach tank circuits with superconducting NbN inductors and internal quality factors $Q_{\mathrm{i}}>1000$ to the interdot barrier gate of silicon double quantum dot devices. Measuring the interdot transition in transport, we quantify radio-frequency crosstalk that results in a ring-up of the resonator when neighbouring plunger gates are driven with frequency components matching the resonator frequency. This effect complicates qubit operation and scales with the loaded quality factor of the resonator, the mutual capacitance between device gate electrodes, and with the inverse of the parasitic capacitance to ground. Setting qubit frequencies below the resonator frequency is expected to substantially suppress this type of crosstalk.

\end{abstract}

\maketitle
High-bandwidth readout of spin qubits can be achieved by radio-frequency (RF) reflectometry~\cite{vigneau_probing_2023}, where an RF signal is reflected off a resonator that is either connected directly to a gate of the quantum dot (QD) that defines the spin qubit, or to additional QDs that serve as charge sensors~\cite{barthel_fast_2010, house_high-sensitivity_2016}. The former approach is known as gate-based sensing and avoids the additional footprint of the charge sensors and the necessary leads connected to them~\cite{colless_dispersive_2013, pakkiam_single-shot_2018, cripp_gate-reflectometry_2019, urdampilleta_gate-based_2019, zheng_rapid_2019}.  Rather than detecting the absolute charge state of the spin qubit system, this method detects charge susceptibility in the form of a quantum capacitance~\cite{mizuta_quantum_2017}. Pauli spin blockade leads to a spin-dependent tunneling between two neighbouring QDs, which is seen as a variation in the resonator load capacitance~\cite{petersson_charge_2010} and thereby enables the readout of spin states.

The sensitivity of gate-based dispersive readout can be improved by increasing the internal quality factor $Q_\textrm{i}$ and reducing the parasitic capacitance $C_\textrm{p}$ of the resonator~\cite{ahmed_radio-frequency_2018}. Both can be achieved by using a superconducting inductor fabricated from a thin film of a high-kinetic inductance material such as NbN, which also enables a small resonator footprint and is compatible with the magnetic fields necessary for spin qubit operation~\cite{samkharadze_high-kinetic-inductance_2016, niepce_high_2019, zheng_rapid_2019, yu_magnetic_2021, ibberson_large_2021}.

In this work, we show that attaching a high-quality factor resonator to the gate of a spin qubit device drastically increases the sensitivity of that gate to crosstalk with control pulses applied to neighbouring gates, e.g., to manipulate the spin state via electric-dipole spin resonance (EDSR)~\cite{camenzind_hole_2021, koppens_driven_2006, nadj-perge_spinorbit_2010, maurand_cmos_2016, watzinger_germanium_2018, hendrickx_fast_2020, froning_ultrafast_2021}.  We introduce a method of quantifying such AC crosstalk in a dispersive readout setup and apply it to a double QD in a Si fin field-effect transistor (finFET) device with a tank circuit connected to the barrier gate. The tank circuit is composed of a high-kinetic-inductance NbN nanowire, providing a high $Q_{\mathrm{i}}\approx 1500$ and a low $C_{\mathrm{p}}$. The resonator is excited whenever control pulses on neighbouring gate lines spectrally overlap with its resonance frequency, giving rise to a strongly amplified modulation of the barrier gate and thereby of the double-QD confinement potential. The amplitude of the crosstalk voltage induced on the barrier gate is measured in transport by analysing the corresponding broadening of an interdot charge transition line. This provides an efficient way of characterising AC crosstalk on the device level that does not rely on the tune-up and calibration of qubits \cite{xue_benchmarking_2019, Undseth2023, Weinstein2023, Lawrie2023}. In addition to unintentional driving of neighbouring qubits~\cite{Heinz2021}, this ring-up is expected to lead to increased qubit decoherence in systems with strong spin-orbit interaction (SOI), intrinsic to holes in Si~\cite{maurand_cmos_2016, camenzind_hole_2021} and Ge~\cite{hendrickx_single-hole_2020, froning_ultrafast_2021}, which possess highly anisotropic electric-field dependent $g$-tensors~\cite{malkoc_charge-noise_2022, piot_single_2022, hendrickx_sweet-spot_2023}.

We find that in our device, the main contribution to this type of crosstalk comes from capacitance between the bondpads of neighbouring gate electrodes. Our electrical circuit model predicts that the crosstalk scales proportionally with the loaded quality factor $Q_{\mathrm{l}}$ and with the ratio between the crosstalk capacitance $C_{\mathrm{ct}}$ and $C_{\mathrm{p}}$. Above the resonator frequency $f_\textrm{r}$, the crosstalk induced on the barrier gate saturates at a value of $C_{\mathrm{ct}}/C_{\mathrm{p}}$, whereas for frequencies below $f_{\mathrm{r}}$ it is suppressed.  These findings can aid in the the design of spin qubit architectures with gate-based readout.

The QD devices consist of a fin patterned from bulk silicon, along with two gate layers each consisting of a silicon-oxide dielectric and a TiN gate metal patterned in a self-aligned process \cite{Geyer_2021}. For the first device (device A), a double QD is formed by accumulating holes underneath gates P1 and P2 in the second gate layer (GL2), while the tunnel coupling between the dots is tuned by the barrier gate B in the first gate layer (GL1), see Fig.\,{\ref{fig:fig1}}(a). Lead gates LL and LR (GL1) are used to accumulate charge reservoirs which are tunnel coupled to the respective dots and contacted to source and drain contacts made of PtSi. All gates and the contacts are connected to tungsten bondpads through vias in a silicon oxide encapsulation layer. Devices similar to the ones used here have been shown to host hole spin qubits with operation of both single- and two-qubit gates~\cite{camenzind_hole_2021, geyer_two-qubit_2022}. 

High-kinetic-inductance superconducting nanowire inductors with a wire width of 400\,nm  are fabricated by dry etching a $\sim$\SI{12}{\nano\meter} thick film of NbN exhibiting a nominal sheet inductance of 66\,pH per square, deposited on an intrinsic silicon substrate by DC magnetron sputtering.  A scanning electron microscopy image of a typical inductor is shown in Fig.\,{\ref{fig:fig1}}(d). One end of such an inductor with a nominal inductance of \SI{1.5}{\micro\henry} was connected to the barrier gate of device A by wirebonding from the inductor chiplet to the QD chiplet. The other end of the inductor was connected to a (multiplexed) readout line on the PCB [Fig.\,{\ref{fig:fig1}}(a)]. Such multi-module assemblies consisting of a resonator chiplet separate to the spin qubit device chiplet offer advantages in terms of separation of fabrication steps and choice of materials~\cite{holman_3d_2021,ibberson_large_2021, vonHorstig_multi-module_2023}.

The resonator formed by this inductor together with $C_{\mathrm{p}}$ has a resonance frequency $f_\textrm{r}$ of $\sim$\SI{299}{\mega\hertz}.  The magnitude and phase response is plotted in Fig.\,{\ref{fig:fig1}(b). The superconducting nature of the inductor leads to a large $Q_\textrm{i}$ of $1480 \pm 480$, as determined from a fit of the resonance circle in the complex plane with the method outlined in \cite{probst_efficient_2015}. The large error bar in $Q_{\mathrm{i}}$ arises from the resonator being overcoupled, with a loaded quality factor of $Q_\textrm{l} = 370\pm 50$ that is dominated by the coupling quality factor of $Q_\textrm{c} = 500 \pm 56$. Modeling the parasitic capacitance as a lumped element attached to the devices side of the inductor provides an estimate of $C_{\mathrm{p}}=$ \SI{0.19}{\pico\farad}. The resonance frequency does not exactly match the point at which the magnitude of the reflection coefficient $|S_{11}|$ is minimal as displayed in Fig.\,{\ref{fig:fig1}(b). This is  due to a rotation of the resonance circle in the complex plane, typically attributed to non-ideal interference effects~\cite{khalil_analysis_2012}.

\begin{figure}[!htp]
\includegraphics[width=\linewidth]{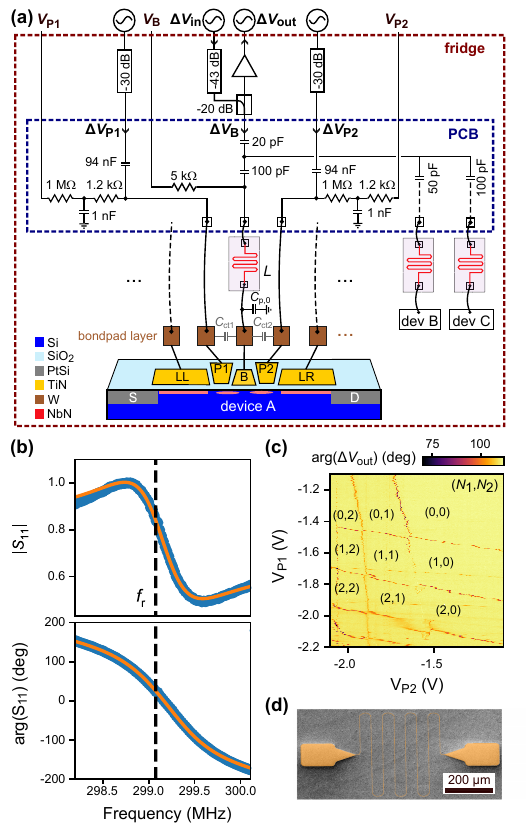}
\caption{ \textbf{(a)} Reflectometry setup with a NbN nanowire inductor on a separate Si chiplet, wire bonded on one side to the barrier gate of a Si finFET double QD device, and on the other side to a multiplexed readout line on a printed circuit board (PCB). Total parasitic capacitance $C_{\mathrm{p}}$ is given by capacitance to ground $C_{\mathrm{p},0}$ and sum of crosstalk capacitances $C_{\mathrm{ct}i}$. \textbf{(b)}  Normalised reflection amplitude and phase of the resonator and corresponding fit, giving $Q_\textrm{i} = 1478 \pm 480$ and $Q_\textrm{c} = 500 \pm 56$. \textbf{(c)} Charge stability diagram of the finFET double-QD at fixed $V_{\mathrm{B}} = -0.845\,\textrm{V}$ obtained by reflectometry at \SI{299.3}{\mega \hertz} revealing the few-hole regime with $(N_1,N_2)$ holes in the two QDs.  \textbf{(d)} False-coloured scanning electron microscope image of a similar NbN nanowire inductor with a wire width of 400\,nm. }
\label{fig:fig1}
\end{figure}

\begin{figure}[!htp]
\includegraphics[width=\linewidth]{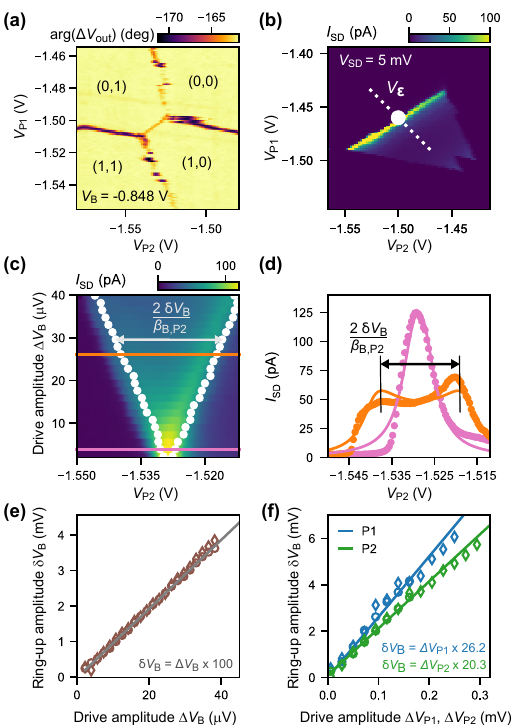}
\caption{\textbf{(a)} Charge stability diagram at the $(0,1)-(1,0)$ transition, detected in the phase of $V_{\mathrm{out}}$ by indirect excitation on gate P1. \textbf{(b)} Finite bias triangles measured in $I_{\mathrm{SD}}$ at the $(0,1)-(1,0)$ transition, and axis of the detuning voltage $V_{\varepsilon}$.  \textbf{(c)} $I_{\mathrm{SD}}$ as a function of $V_{\mathrm{P2}}$ while varying the AC drive amplitude $\Delta V_{\mathrm{B}}$ applied to gate B at frequency $f_{\textrm{r}}$. The peak broadening (white dots, arrow shows twice the broadening amplitude) is a measure of the resonator ring-up amplitude on gate B. \textbf{(d)} Linecuts  of $I_{\mathrm{SD}}$ for two different sinusoidal drive amplitudes $\Delta V_\mathrm{B}$ and their corresponding fits. \textbf{(e)} Extracted ring-up amplitude $\delta V_{\mathrm{B}}$ on gate B as a function of $\Delta V_{\mathrm{B}}$, yielding an average amplification of $\Delta V_{\mathrm{B}}$ by a factor of $100\pm3$ (solid line). \textbf{(f)} $\delta V_{\mathrm{B}}$ as a function of AC drive amplitudes $\Delta V_{\mathrm{P1}}$ ($\Delta V_{\mathrm{P2}}$) applied to neighbouring gates P1 (P2), demonstrating amplification of the drive signal by factors 26.2 (20.3). In (e) and (f), diamonds (circles) represent data obtained for DC scans of the interdot lines along P1 (P2). }
\label{fig:fig2}
\end{figure}

The resonator can be operated as a gate-based dispersive sensor by probing the reflected signal at resonance while sweeping the plunger gates. The obtained charge stability diagram of the double QD system is shown in Fig.\,\ref{fig:fig1}(c) with clear dot-to-lead transitions visible down to the last hole. Interdot transitions are also visible because gate B has different lever arms to the two dots. The signal amplitude is smaller than that of the dot-to-lead transitions.

Surprisingly, the charge stability diagram can also be observed in $\Delta V_{\mathrm{out}}$ when driving a neighbouring plunger gate. Such an indirect excitation of the resonator indicates the presence of a finite crosstalk capacitance $C_{\textrm{ct}}$ between that plunger gate and gate B. The measured phase of $\Delta V_{\mathrm{out}}$ when exciting an AC amplitude $\Delta V_{\mathrm{P1}}$
on gate P1 at resonance is shown in Fig.~\ref{fig:fig2}(a) around the  $(0,1)-(1,0)$ charge region. We note that such crosstalk is distinct from the harmonic voltage conversion observed in Ref.~\cite{oakes_quantum_2022}, where it is induced by transitions of single charges at interdot and dot-to-lead transitions. 

The crosstalk in this device is quantified by measuring the broadening of the interdot charge transition line in the source-drain current $I_{\mathrm{SD}}$ and relating this to a ring-up peak voltage amplitude $\delta V_{\mathrm{B}}$ on gate B. We first determine the line broadening when directly exciting the tank circuit with an oscillation amplitude $\Delta V_{\mathrm{B}}$ [see Fig.\ref{fig:fig1}(a)] at a frequency that is on resonance with the tank circuit. The typical bias triangles with a bright interdot line are shown in a map of  $I_{\mathrm{SD}}$ in Fig.\,\ref{fig:fig2}(b) for an applied source-drain bias $V_{\textrm{SD}}$ of \SI{5}{\milli\volt} and with $\Delta V_{\mathrm{B}}=0$. Fig.\,\ref{fig:fig2}(c) shows scans of the interdot line along the DC voltage $V_{\mathrm{P2}}$ and for different amplitudes $\Delta V_{\mathrm{B}}$. The broadening is fit by a time-averaged sinusoidally-shifted Lorentzian function, see Fig.\,\ref{fig:fig2}(d) and supplementary material Sec.\,I for details. The amplitude of the broadening (i.e., half the distance between the extreme positions of the fitted Lorentzian peaks) is more than 400 times larger than $\Delta V_{\mathrm{B}}$ and is a consequence of a resonant ring-up of the voltage on gate B. The amplitude $\delta V_{\mathrm{B}}$ of this ring-up is obtained by multiplying the broadening amplitude scanned along $V_{\mathrm{P}i}$ by $\beta_{\mathrm{B,P}i}$, where $\beta_{\mathrm{B,P}i}$ denotes the ratio between voltage changes on gate B and on gate $\mathrm{P}i$ required to stay on the interdot line. We find $\beta_{\mathrm{B,P2}}=-0.23$ and  $\beta_{\mathrm{B,P1}}=0.40$, see Sec.\,II of the supplementary material.

Figure \ref{fig:fig2}(e) shows the obtained amplitude $\delta V_{\mathrm{B}}$ as a function of $\Delta V_{\mathrm{B}}$ by individually fitting the scans of the interdot line along P1 and P2 and adjusting for the relative voltage ratios $\beta_{\mathrm{B,P1}}$ and $\beta_{\mathrm{B,P2}}$ respectively. We find a linear relationship with an average amplification factor of $100\pm3$. This value is consistent with the amplification factor as calculated from numerical simulation of the readout circuit (see Sec.\,IV of the supplementary material).

Using the same method but for indirect excitation of the tank circuit by a resonant AC drive with amplitude $\Delta V_{\mathrm{P1}}$ ($\Delta V_{\mathrm{P2}}$) on gate P1 (P2), we find a significant ring-up of gate B with an amplitude $\delta V_{\mathrm{B}}$ that is 26.2 (20.3) times larger than the exciting amplitude on the P1 (P2) gate [Fig.\,\ref{fig:fig2}(f)]. As we show next, this crosstalk-induced excitation of the resonator and thereby of the potential on gate B occurs for any signal that contains spectral components within the bandwidth of the resonator frequency.

\begin{figure*}[!htp]
\includegraphics[width=\textwidth]{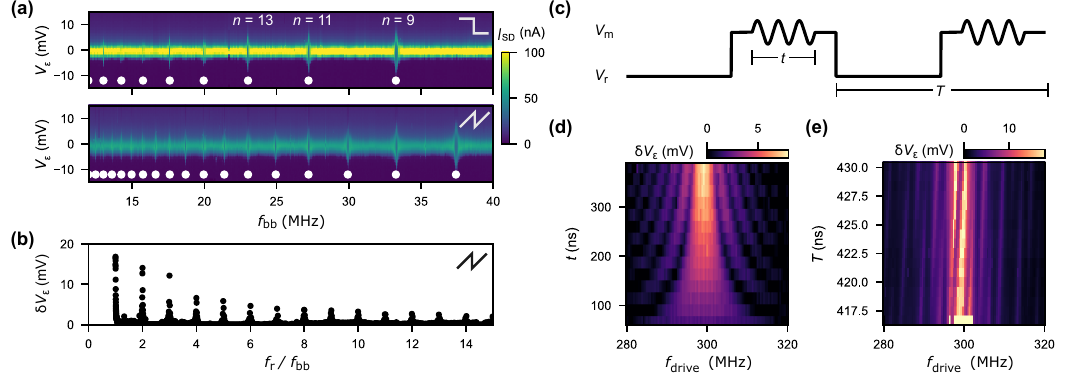}
\caption{ \textbf{(a)} Measurement of $I_{\mathrm{SD}}$ while applying a square (upper) and sawtooth (lower) wave with frequency $f_{\mathrm{bb}}$ and peak amplitude at gate P1 of \SI{3.15}{mV}.  \textbf{(b)} Fitted peak broadening $\delta V_{\varepsilon}$, revealing the amplitudes of the harmonics $n=f_{\mathrm{r}} / f_{\mathrm{bb}}$ of the sawtooth wave applied. \textbf{(c)}  Typical Rabi pulse sequence with a square wave component (period $T$) alternating between the manipulation point $V_{\textrm{m}}$ and the readout point $V_{\textrm{r}}$, superposed with a sinusoidal qubit drive pulse of length $t$. \textbf{(d)} Peak broadening $\delta V_{\varepsilon}$ when Rabi pulses with frequency $f_{\mathrm{drive}}$ and length $t$ are applied repeatedly with period $T=$\SI{1}{\micro\second} and peak amplitude of \SI{0.7}{\milli\volt} at gate P1. \textbf{(e)} Same as (d) but varying the baseband pulse period $T$.}
\label{fig:fig3}
\end{figure*}

In spin qubit experiments, baseband signals are typically applied to plunger gates when transitioning from a qubit manipulation point to a readout point in charge configuration space. The repetition of such baseband signals may lead to harmonics that excite the resonator gate. To illustrate this effect, we apply a square wave [Fig.\,\ref{fig:fig3}(a), upper] or sawtooth wave [Fig.\,\ref{fig:fig3}(a), lower] of varying frequency $f_{\mathrm{bb}}$ to gate P1. 
We fit the broadening of the interdot line $\delta V_\varepsilon$ by sweeping the interdot detuning voltage $V_\varepsilon$ (see Sec.\,III of the supplementary material for definition) across the $(0,1)-(1,0)$ transition, as indicated in Fig.\,\ref{fig:fig2}(b). When varying the baseband frequency $f_{\mathrm{bb}}$, a crosstalk-induced broadening of the interdot line is observed every time a harmonic $n$ of the signal matches the resonator frequency $f_{\mathrm{r}}$, see Fig.~\ref{fig:fig3}(a). Figure \ref{fig:fig3}(b) shows the fitted interdot peak broadening $\delta V_{\varepsilon}$ as a function of $f_{\mathrm{r}}/f_{\mathrm{bb}}$ for a sawtooth wave. Excitations occur whenever $f_{\mathrm{bb}}\cross n=f_\textrm{r}$, indicated by the white dots in Fig.~\ref{fig:fig3}(a), with the expected Fourier amplitudes scaling with $\frac{1}{n}$. Similarly, only odd harmonics are observed for a square wave excitation. The adverse effect of this ring-up  on the operation of QDs as qubits can be reduced for larger $n$ by filtering the baseband pulses. However, fast ramp times between different charge states may be necessary to fulfill diabaticity requirements when initialising spin states via rapid adiabatic passage~\cite{taylor_relaxation_2007, harvey-collard_spin-orbit_2019}. 

When manipulating a spin qubit, a series of sinusoidal drive pulses are applied to the gates, see Fig.\,\ref{fig:fig3}(c). We focus on a typical Rabi experiment where the duration $t$ of the drive pulses is varied in order to observe Rabi oscillations induced by EDSR. The fitted interdot peak broadening for such a pulse train with a repetition rate 1/$T$ of \SI{1}{\mega \hertz} is shown in Fig.\,\ref{fig:fig3}(d), where both the drive pulse frequency $f_{\mathrm{drive}}$ and the pulse duration $t$ are swept. The baseband amplitude is set to zero. The observed peak broadening as a function of $f_{\mathrm{drive}}$ and $t$ matches the sinc function $\sin(x)/x$ expected for the Fourier transformation of a sinusoidal pulse of a finite length, with $x=\pi t (f_{\mathrm{drive}}-f_{\mathrm{r}})$. Note that the observed pattern resembles but is not related to the typical Rabi chevron pattern observed when varying $f_{\mathrm{drive}}$ and $t$ for pulses applied to a qubit. While the fringes of the sinc function can be reduced by using a Gaussian envelope of the drive pulses, the broadening of the Fourier spectrum remains and therefore extends the crosstalk into a bandwidth $1/t$ around $f_\mathrm{r}$. We additionally resolve the effect of the repetition of the pulses with period $T$ as lines spaced by the repetition rate, as displayed in  Fig.\,\ref{fig:fig3}(f) for a fixed $t=\SI{250}{\nano\second}$.

\begin{figure}[]
\includegraphics{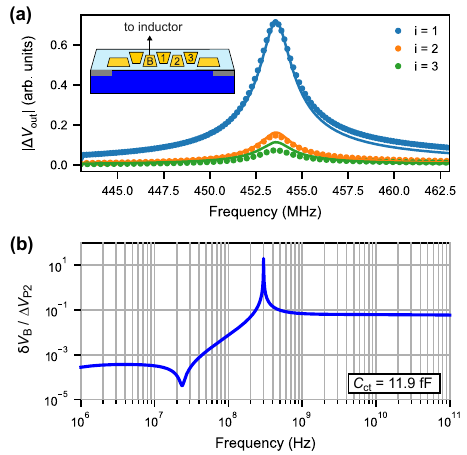}
\caption{\textbf{(a)} Measured signal $|\Delta V_{\mathrm{out}}|$ (dots) transmitted through resonator attached to gate B of device B (inset) when applying an AC drive to one of the neighbouring gates $i$. The simulated transmitted signal $|\Delta V_{\mathrm{out}}|$ is indicated as solid lines.  \textbf{(b)} Simulated amplification factor $\delta V_{\mathrm{B}}$/$\Delta V_{\mathrm{P2}}$ of the crosstalk as a function of the drive frequency on the neighbouring gate P2.}
\label{fig:fig4_single_col}
\end{figure}

The indirect excitation of the resonator can be reproduced in a discrete-element circuit model, see Sec.\,V of the supplementary material for details of the circuit diagram.  The resonator is indirectly excited by a drive signal on the neighbouring gate through a mutual capacitance $C_{\mathrm{ct}}$. The total parasitic capacitance to ground, $C_{\mathrm{p}}$, is given by the sum of $C_{\mathrm{p},0}$ and $C_{\mathrm{ct}}$, which is kept constant at \SI{0.19}{\pico \farad}. The amplification factors between $\delta V_{\mathrm{B}}$ and $\Delta V_{\mathrm{P1}}$ ($\Delta V_{\mathrm{P2}}$) are found to match those in Fig.\,\ref{fig:fig2}(f) by choosing $C_{\mathrm{ct}}$ equal to \SI{16.0}{\femto\farad} (\SI{11.9}{\femto\farad}) when exciting on P1 (P2). This can mostly be accounted for by the mutual capacitance between the corresponding bondpads, for which we find values of \SI{12.1}{\femto\farad} (\SI{7.3}{\femto\farad}) using Ansys Maxwell (Table\,\ref{table:capacitance_matrix}). We assign the remaining coupling capacitance of around \SI{4}{\femto\farad} to mutual capacitance between bond wires and to the PCB side of our setup. Capacitances between neighbouring gates at the device level were found to be on the order of $\sim$\,\SI{0.5}{\femto \farad} and therefore negligible. 

To further confirm that the observed crosstalk is dominated by the capacitance between bondpads, another Si finFET device (device B) was measured. This device has two nanogates in GL1 and three nanogates in GL2, as depicted in the inset of Fig.\,\ref{fig:fig4_single_col}(b), with a \SI{750}{\nano \henry} inductor attached to one of the first-gate-layer gates, named gate B. This leads to the formation of a resonator with a resonance frequency of \SI{454}{\mega \hertz}. Three of the other gates, labelled gate 1 (nearest neighbour), gate 2 (next-nearest neighbour), and gate 3 (next-next nearest neighbour), were individually excited with an AC drive tone of varying frequency, and the voltage amplitude $|\Delta V_{\mathrm{out}}|$ transmitted through the resonator was measured. The transmitted signal peaks at the resonator frequency. The peak amplitude is a measure for the voltage amplitude at gate B induced by AC crosstalk. A discrete-element circuit model (see Sec.\,V of the supplementary material) was used to model the results. In this model, the various crosstalk capacitances between the different gates were obtained from  electrostatic simulations, taking into account that the bondpad layout where the order of the bondpads is the same as that of the nanogates. The  measured $|\Delta V_{\mathrm{out}}|$ is presented in Fig.\,\ref{fig:fig4_single_col}(a) along with the simulated results from the circuit model (solid lines). The transmission magnitude decreases with the distance of the excited gate to gate B, in good agreement with the simulation. We thus attribute the dominant source of crosstalk to the capacitance between gate electrodes in the bondpad layer. 

The frequency dependence of the AC crosstalk $\delta V_{\mathrm{B}}/\Delta V_{\mathrm{P2}}$ is simulated for device A, see Fig.\,\ref{fig:fig4_single_col}(b) for the case of $C_{\mathrm{ct}} = \SI{11.9}{\femto \farad}$. The maximum value of 20.3 is reached at $f_{\mathrm{r}}$. For our typical case, $C_{\mathrm{p}}\gg C_{\mathrm{ct}}$, this maximum value is well approximated by $Q_{\mathrm{l}} C_{\mathrm{ct}} / C_{\mathrm{p}}$. Below $f_{\mathrm{r}}$, the crosstalk reaches a minimum and is suppressed. Above $f_{\mathrm{r}}$, it saturates at a value of approximately $C_{\mathrm{ct}}/C_{\mathrm{p}}=6\%$ . This suggests that placing qubit frequencies well below the resonator frequency is optimal to suppress crosstalk in architectures with gate-based dispersive qubit readout. Note that our model does not account for higher-order resonator modes, where additional resonances at higher frequencies would lead to crosstalk amplitudes well above this saturation value. 

\begin{table}[]
\begin{tabular}{cccc}
            & \textbf{P1} & \textbf{B} & \textbf{P2} \\ \hline
\textbf{P1} & \SI{51.8}{fF}    & \SI{12.1}{fF} & \SI{1.7}{fF}    \\
\textbf{B}  & \SI{12.1}{fF}  & \SI{42.0}{fF}  & \SI{7.3}{fF}    \\
\textbf{P2} & \SI{1.7}{fF} & \SI{7.3}{fF} & \SI{41.9}{fF}    \\ \hline
\end{tabular}
\caption{\label{table:capacitance_matrix} Simulated  capacitance matrix between the bondpads of the gate electrodes for device A.}
\end{table}

In conclusion, gate-based dispersive sensing has been used to measure the charge stability diagram of a Si finFET double QD device down to the last hole with a superconducting NbN tank circuit attached to the barrier gate. The high quality factor of the resonator comes at the expense of increased AC crosstalk where a ring-up of the gate electrode attached to the tank circuit is observed when  other gate electrodes are driven with frequency components matching the resonator frequency. We have demonstrated a method of quantifying such crosstalk by measuring the source-drain current through the double QD device and fitting the broadening of the interdot current peak due to the ring-up of the resonator gate. 

These findings identify some limitations of gate-based dispersive sensing as a qubit readout technique. The crosstalk amplitude on resonance scales as $Q_{\mathrm{l}}C_{\mathrm{ct}}/C_{\mathrm{p}}$ and reaches the limit $C_{\mathrm{ct}}/C_{\mathrm{p}}$ above the resonator frequency $f_{\mathrm{r}}$. Pulses applied to QD gates that contain spectral components above $f_{\mathrm{r}}$ should therefore be avoided, even more so if higher modes of the resonator exist. Below the resonator frequency, crosstalk is found to be largely suppressed. This suggests to place qubit frequencies well below $f_{\mathrm{r}}$ to increase the bandwidth for qubit driving.
In general, crosstalk can be reduced by optimising the bondpad layout, improving signal routing or using differential signalling schemes \cite{Blanvillain2012, Machida2023, Heinz2021}. Unfortunately, an increase of $C_{\mathrm{p}}$ or decrease of $Q_{\mathrm{l}}$, while reducing the crosstalk, decreases the sensitivity of the readout circuit to the device capacitance, i.e., the spin readout signal. One way to overcome this would be to reduce the interaction of the spin qubits with the resonator in the qubit manipulation phase, e.g. using a transistor between tank circuit and QD gate~\cite{schaal_conditional_2018} or by switching $C_{\mathrm{p}}$ by varactor diodes. 

\vspace{5mm} 
See supplementary material for details on the fit function for the broadening of the interdot peak, measured ratios of voltage changes on each gate to stay on the interdot line, definition of detuning voltage, circuit models of both devices and the device bondpad layout.

\vspace{5mm} 

This project has received funding from the European Union’s Horizon 2020 research and innovation programme under the Marie Skłodowska-Curie grant agreement number 847471, and by the NCCR SPIN under grant number 51NF40-180604 of the Swiss National Science Foundation. We thank the Cleanroom Operations Team of the Binnig and Rohrer Nanotechnology
Center (BRNC) for their help and support.

%

\end{document}


\title{Supplementary Material: Capacitive crosstalk in gate-based dispersive sensing of spin qubits}

\author{Eoin G. Kelly}
\author{Alexei Orekhov}
\author{Nico Hendrickx}
\author{Matthias Mergenthaler}
\author{Felix Schupp}
\author{Stephan Paredes}
\affiliation{IBM Research -- Zurich, Säumerstrasse 4, CH-8803 Rüschlikon, Switzerland}
\author{Rafael S. Eggli}
\author{Andreas V. Kuhlmann}
\affiliation{Department of Physics, University of Basel, Klingelbergstrasse 82, CH-4056 Basel, Switzerland}
\author{Patrick Harvey-Collard}
\author{Andreas Fuhrer}
\author{Gian Salis}
\affiliation{IBM Research -- Zurich, Säumerstrasse 4, CH-8803 Rüschlikon, Switzerland}
\email{gsa@zurich.ibm.com}

\date{\today}

\maketitle

\section{Interdot Peak Broadening Fit Function}
\label{app:interdot_fit}

The current $I$ of an interdot peak broadened by an applied AC drive $\Delta V_{\mathrm{d}}\sin(x)$ is fit with a time-averaged sinusoidally-broadened Lorentzian function:

\begin{equation}
I(V_{\mathrm{P}i}) = \frac{I_0\gamma^2}{2\pi}\int^{2\pi}_{0}\frac{1}{(V_{\mathrm{P}i}-\Delta V_{\mathrm{d}}\sin x - V_0)^2 + \gamma^2}dx
\end{equation}

Here $V_{\mathrm{P}i}$ is the voltage applied to a gate $\mathrm{P}i$, $\gamma$ is the interdot peak linewidth, $V_0$ the peak position in voltage and $I_0$ the peak current for $\Delta V_{\mathrm{d}}=0$.

\section{Measurement of relative lever arms with respect to interdot line}
\label{app:rel_lever_arms}

\begin{figure}[H]
\includegraphics{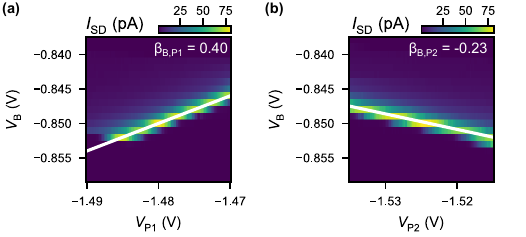}
\caption{Extraction of relative voltage ratios $\beta_{\mathrm{B,P1}}$ and $\beta_{\mathrm{B,P2}}$ for a change in $V_{\mathrm{B}}$ versus a change in \textbf{(a)} $V_{\mathrm{P1}}$ and \textbf{(b)}  $V_{\mathrm{P2}}$ along the $(1,0)-(0,1)$ interdot line, with the obtained values shown in the inset.}
\label{fig:rel_ever_arms}
\end{figure}

\section{Detuning Voltage}
\label{app:detuning_voltage}
The detuning voltage $V_{\mathrm{\varepsilon}}$ is here defined as $V_{\mathrm{P1}}=V_{\mathrm{P1},0}-V_{\mathrm{\varepsilon}}$ and $V_{\mathrm{P2}}=V_{\mathrm{P2},0}+V_{\mathrm{\varepsilon}}$, with $V_{\mathrm{\varepsilon}}=0$ located on the interdot line where $V_{\mathrm{P1}}=V_{\mathrm{P1},0}$ and $V_{\mathrm{P2}}=V_{\mathrm{P2},0}$.

\section{Circuit Model  of Device A}\label{app:circuit_threegate}

\begin{figure}[H]
\includegraphics{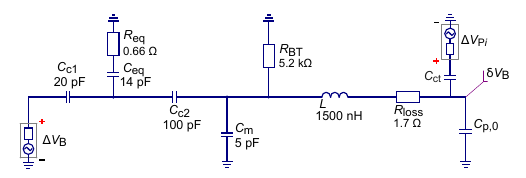}
\caption{ Schematic the circuit used to simulate the direct excitation of the resonator and crosstalk-induced excitation between the different gates of device A.}
\label{fig:fig5}
\end{figure}

The circuit used to simulate the response of the device A is shown in Figure~\ref{fig:fig5}. The resonator consists of an inductor $L$, a parasitic capacitance to ground $C_{\mathrm{p},0}$, and a resistor $R_{\mathrm{loss}}$ introduced to model the resonator damping. A bias is applied to the inductor via the resistor $R_{\mathrm{BT}}$. The resonator is coupled to a multiplexed line through a coupling capacitor $C_{\mathrm{c2}}$, where the other resonators on the multiplexed line have been substituted by an equivalent capacitor $C_{\mathrm{eq}}$ and resistor $R_{\mathrm{eq}}$ at the resonator frequency. The resonator can be directly excited by a drive $\Delta V_{\mathrm{B}}$ at the multiplexed coupling capacitor $C_{\mathrm{c1}}$. The capacitance to ground $C_{\textrm{m}}=$ \SI{5}{\pico\farad} is intrinsic to the PCB. The resonator is indirectly excited by a drive on the neighbouring gate via a crosstalk capacitance $C_{\textrm{ct}}.$

\begin{figure}[H]
\includegraphics{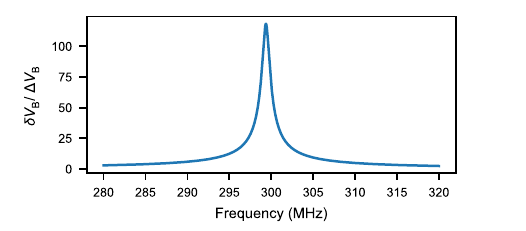}
\caption{Simulated frequency response of the directly excited readout circuit for device A.}
\label{fig:fig_app3}
\end{figure}

The simulated amplification factor as a function of the excitation frequency for a direct excitation of device A with $C_{\mathrm{ct}}+C_{\mathrm{p},0}=$\SI{0.19}{\pico\farad} is displayed in Fig.\,\ref{fig:fig_app3}. The amplification factor on resonance is found to be 119.

\section{Circuit Model of Device B} \label{app:circuit_fivegate}

\begin{figure}[H]
\includegraphics[width=\linewidth]{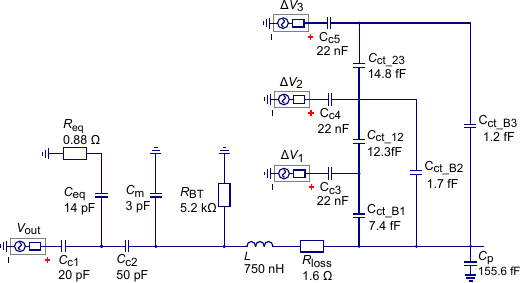}
\caption{ Schematic the circuit used to simulate the crosstalk between the different gates of device B. }
\label{fig:fig6}
\end{figure}

\begin{figure}[H]
\centering
\includegraphics[]{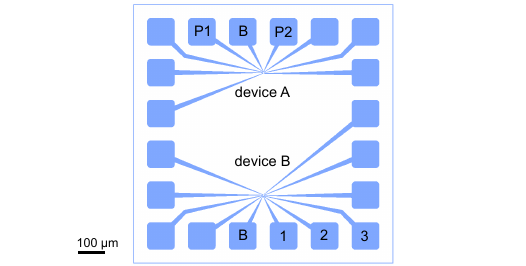}
\caption{Bondpad layout for device A (top) and device B (bottom).}
\label{fig:fig7}
\end{figure}

The circuit used to simulate the response of device B is shown in Fig.\,\ref{fig:fig6}, where the multiplexed readout line has been replaced by an equivalent capacitance and resistance to ground at the resonance frequency of \SI{454}{\mega\hertz}.  The capacitances $C_{\mathrm{ct\_ij}}$ between gates $i$ and $j$ were found from electrostatic simulations of the bondpad layout (see Fig.\ref{fig:fig7}) using Ansys Maxwell. The simulation assumes a \SI{150}{\nano\meter} metallization layer of tungsten on top of \SI{525}{\micro\meter} of Si with a relative permittivity of 11.9 and a vacuum padding of \SI{200}{\micro\meter} on all sides. A substrate area of 1100$\times$\SI{1100}{\micro\meter} is used. The bottom of the chip is set to a ground potential. 
\\
